\documentclass[a4paper,conference]{IEEEtran}
\usepackage{cite}

\usepackage{comment}
\newcommand{\bm}[1]{{\mbox{\boldmath $#1$}}}

\ifCLASSINFOpdf
  \usepackage[pdftex]{graphicx}
  \graphicspath{{../pdf/}{../jpeg/}}
\else
\fi
\usepackage{amsmath,amssymb,amsfonts}
\usepackage{url}

\usepackage{color}
\usepackage{soul}
\usepackage{hyperref}
\usepackage{xcolor}
\hypersetup{
    colorlinks=true,
    urlcolor=blue,
}

\begin{document}
%
\title{Fidelity-Controllable Extreme Image Compression with Generative Adversarial Networks}

\author{
\IEEEauthorblockN{Shoma Iwai, Tomo Miyazaki, Yoshihiro Sugaya, and Shinichiro Omachi}
\IEEEauthorblockA{Tohoku University\\\{shoiwai, tomo, sugaya, machi\}@iic.ecei.tohoku.ac.jp}
}


%


\maketitle

\begin{abstract}
We propose a GAN-based image compression method working at extremely low bitrates below 0.1bpp. Most existing learned image compression methods suffer from blur at extremely low bitrates. Although GAN can help to reconstruct sharp images, there are two drawbacks. First, GAN makes training unstable. Second, the reconstructions often contain unpleasing noise or artifacts.
To address both of the drawbacks, our method adopts two-stage training and network interpolation. The two-stage training is effective to stabilize the training. 
Moreover, the network interpolation utilizes the models in both stages and reduces undesirable noise and artifacts, while maintaining important edges. Hence, we can control the trade-off between perceptual quality and fidelity without re-training models. The experimental results show that our model can reconstruct high quality images. Furthermore, our user study confirms that our reconstructions are preferable to state-of-the-art GAN-based image compression model. Our code will be available at \href{https://github.com/iwa-shi/fidelity_controllable_compression}{\url{https://github.com/iwa-shi/fidelity_controllable_compression}}
\end{abstract}


%
\IEEEpeerreviewmaketitle

\section{Introduction}
Image compression is an important technique for efficient image storage. 
Recently, machine-learning based image compression methods have been studied~\cite{balle17,balle18,minnen,gmm}. Some approaches outperform engineered codecs, such as JPEG, JPEG2000, and BPG~\cite{bpg}.
Generally, machine-learning based methods are trained to minimize the rate-distortion object function in end-to-end fashion. 
\begin{equation}
    \min R + \lambda D \label{rd-tradeoff}
\end{equation}
$R$, $D$ and $\lambda$ are a rate term, a distortion term, and a Lagrange multiplier, respectively. 
$R$ represents the entropy of latent codes, which is estimated by an entropy model. 
$D$ means difference between the original image and the compressed one, e.g., mean squared error (MSE) and multi-scale structural similarity (MS-SSIM).
$\lambda$ determines the desired rate–distortion trade-off. The compression ratio is high when $\lambda$ is small, however, the compressed images suffer from blur. Especially at a very low bitrate, perceptual quality of a reconstruction becomes so poor.

\begin{figure}[t]
    \centering
    \includegraphics[width=\linewidth]{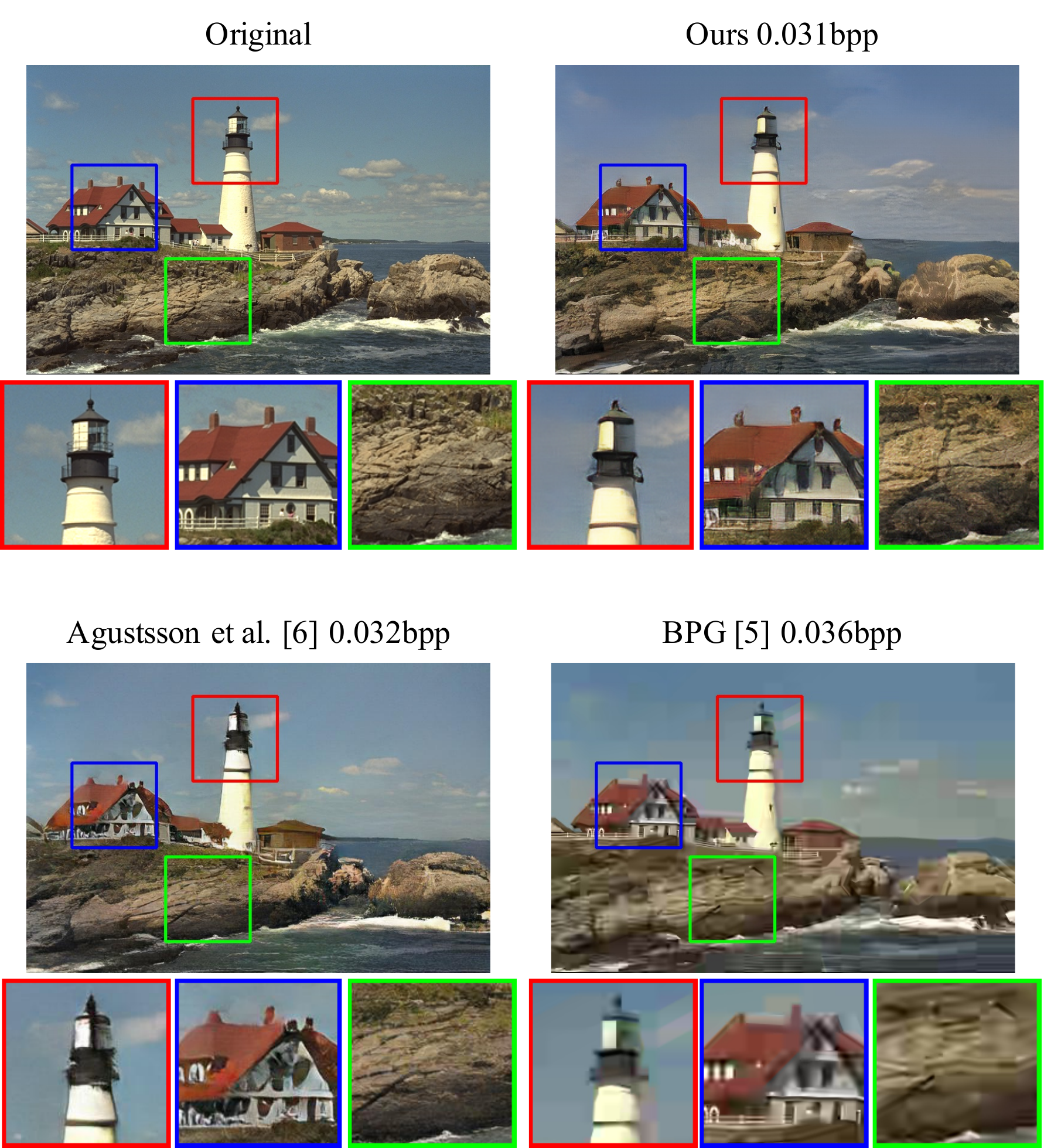}
    \caption{Visual comparison of the original image, our reconstruction, the reconstruction of Agustsson et al.\cite{agustsson} (another GAN-based image compression model) and BPG\cite{bpg} (state-of-the-art engineered image compression codec). Our reconstruction doesn't suffer from collapse or blur.}
    \label{fig:kodim21}
\end{figure}

To tackle the problem, some methods~\cite{rippel,agustsson,a_gan_based,lee_clic} adopt generative adversarial networks (GAN). Although GAN makes distortion high, perceptual quality becomes better. Moreover, they produce less blurry compressed images even at low bitrates. However, there are two main drawbacks in GAN-based methods. First, GAN-based methods suffer from unstable training. Second, reconstructions often contain undesirable noise or artifacts.

In this paper, we focus on the very low bitrate below 0.1 bit per pixel (bpp) compression. We propose a novel GAN-based image compression method that overcomes the aforementioned two problems.
In order to make training stable, we utilize a two-stage training \cite{lee_clic}. 
In the first stage, we train both of the encoder and decoder by optimizing (\ref{rd-tradeoff}) without GAN modules.
In the second stage, we fine-tune only the decoder with GAN.
Since this training strategy makes the optimization simple, training becomes stable. 
Moreover, inspired by recent advance in super resolution \cite{esrgan}, we merge two decoders with/without GAN to suppress noises and artifacts. This enables us to control the balance between perceptual quality and fidelity of reconstructions.
Therefore, our reconstructions can avoid degradation while maintaining high perceptual quality.
Futhermore, inspired by \cite{gmm}, we utilize Gaussian mixture model (GMM) for flexible entropy model. 
\par
As shown in Fig.~\ref{fig:kodim21}, the proposed method yields visually more pleasing results than the engineered codec BPG \cite{bpg} and another GAN-based compression model Agustsson et al.\cite{agustsson} at extreme low bitrates. BPG is blurry and \cite{agustsson} suffers from excessive noise. In contrast, the proposed method can reconstruct a high quality and high fidelity image.

The main contributions of this paper are followings:
\begin{itemize}
\item We combine the compression method \cite{gmm} with GAN and achieve to reconstruct high quality images at extreme low bitrates (below 0.1bpp). 
\item We propose to merge two decoders with or without GAN into one decoder so that we can suppress undesirable noise and artifacts in reconstructions.
\item Our user study verified that the proposed method achieved better visual quality than the existing GAN-based method \cite{agustsson}. 
\end{itemize}

\section{Related Works}
\subsection{Generative Adversarial Networks}
Generative adversarial networks (GAN) \cite{GAN} is an active area of research. GAN has two components, generator $G$ and discriminator $D$. These networks are trained to optimize the adversarial min-max problem:
\begin{equation}
    \min_G \max_D \mathbb{E} [\log D(x)] + \mathbb{E} [\log (1-D(G(z))]
\end{equation}
$x$ and $z$ are a real image and a latent code, respectively. 
The generator learns to fool the discriminator, whereas the discriminator learns to distinguish real and fake images. 
In this way, the generator can generate photo-realistic images. GAN has been used in image restoration tasks such as super-resolution \cite{esrgan}, image deblurring \cite{deblur_gan}, or image-inpainting \cite{deepfillv2} as well as image generation because GAN helps the model to reconstruct realistic images.

\subsection{Learned Image Compression}
Recently, a lot of deep-learning based image compression methods have been proposed\cite{balle17, balle18, minnen, conditional_probability, ca_entropy, gmm, NLAIC}. 
\cite{balle17} proposes end-to-end trainable image compression system which optimize rate-distortion trade-off (\ref{rd-tradeoff}). 
Some works investigate the effective entropy model. \cite{balle18} introduced hyperprior networks that utilize side information to estimate variance parameters of distributions of latent codes. \cite{minnen,ca_entropy} combined a hyperprior with a 2D PixelCNN-based \cite{pixelcnn} context model. These methods estimate the mean and variance of the distribution of latent codes. \cite{conditional_probability} adopts 3D-CNN in the context model and uses an importance map for adaptive bit allocation. \cite{gmm} utilizes the Gaussian mixture model instead of a single Gaussian distribution to achieve a more flexible entropy model. Other works focus on the architectures of networks. \cite{multi_scale_comp} proposes a multi-scale encoder-decoder network. \cite{NLAIC} adopts non-local attention modules to capture both local and global features. Since the methods mentioned above learn to optimize rate-distortion trade-off (\ref{rd-tradeoff}), the reconstructions suffer from blur at low bitrates. \par
Some methods \cite{rippel, agustsson, a_gan_based, generative_model_dp,lee_clic} adopt GAN framework to reconstruct realistic images. \cite{rippel} introduces adversarial training and pyramidal decomposition. However, \cite{rippel} adopts a complex and heuristic training procedure for stable training, and it is a non-standard GAN training. \cite{agustsson, a_gan_based} achieved to reconstruct realistic images at low bitrates. These methods use no rate term in their loss functions for stable training, but it can lead to suboptimal bitrates.
\cite{generative_model_dp} can generate photo-realistic images at every bitrate. However, this method focuses on domain-specific datasets such as face or bedroom images.
\cite{lee_clic} proposes two-stage training for stable training. The compression model can reconstruct high perceptual quality images.
Some methods \cite{rippel,agustsson,a_gan_based,lee_clic} tackle the problem of unstable training.  However, they cannot reduce the undesirable noise or artifacts which may change the impression of the image.
Our method utilizes the two-stage training \cite{lee_clic} and a network-interpolation \cite{esrgan} to tackle both the problems.


\section{Proposed Method}
\begin{figure*}[t]
    \centering
    \includegraphics[width=\linewidth]{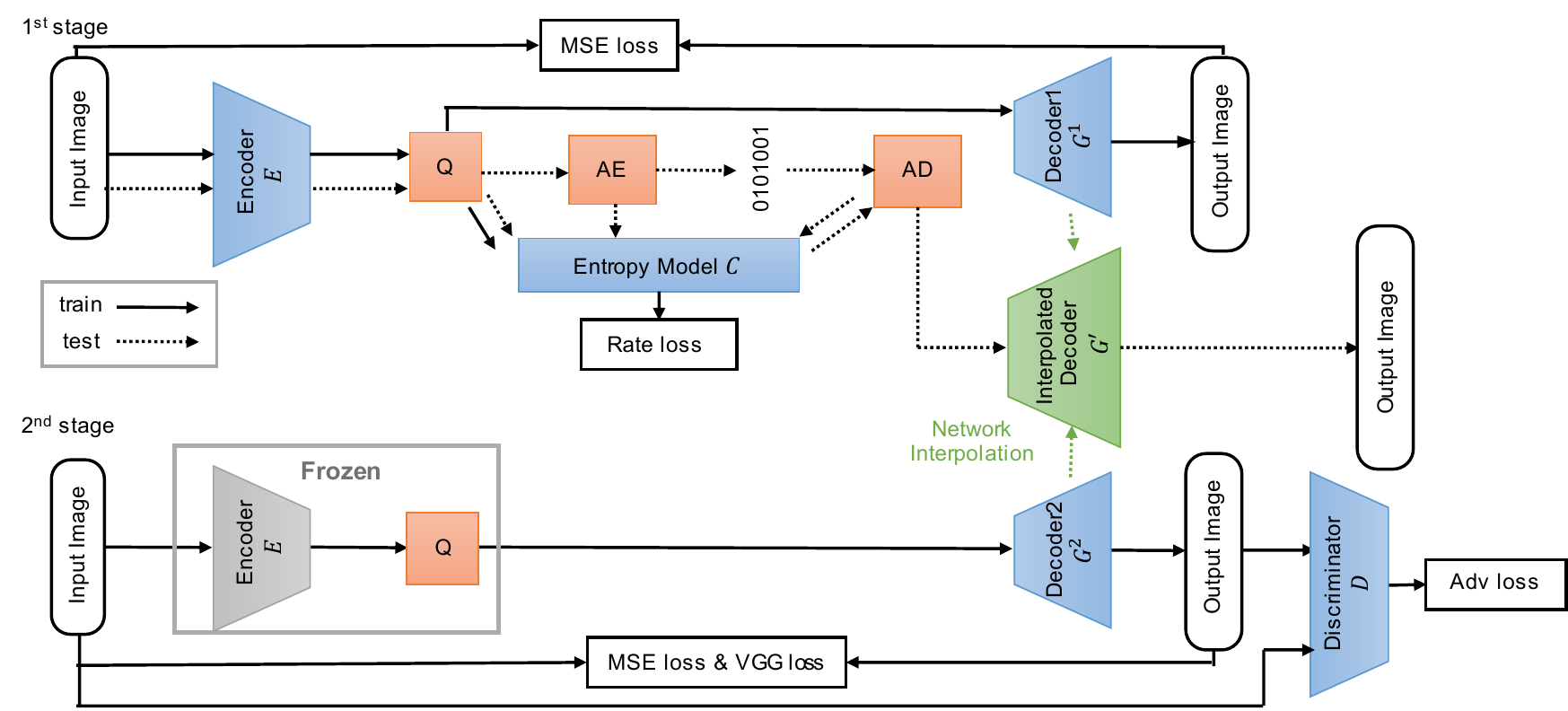}
    \caption{Our overall pipeline. In the training phase (solid arrow), the entire model is trained in the first stage, and the second stage trains only the decoder. In the test phase (dotted arrow), we interpolate a new decorder using the two decoders. The interpolated decoder reconstructs an input image from quantized latent code. Q, AE, and AD are a quantizer, an arithmetic encoder and an arithmetic decoder, respectively.}
    \label{fig:overall}
\end{figure*}
We show the overall pipeline of our compression system in Fig.~\ref{fig:overall}. In this section, we will explain details of the architecture, the two-stage training, and the network interpolation.

\subsection{Image Compression Model}
We developed our image compression model on the basis of the state-of-the-art model \cite{gmm} with two main renovations. First, we add an additional discriminator to adopt adversarial training for improving perceptual quality. Second, we remove hyperprior networks of \cite{gmm} because the model learns to make the entropy of side information zero at extremely low bit rate setting. 

\begin{figure*}[t] \centering
    \includegraphics[width=\linewidth]{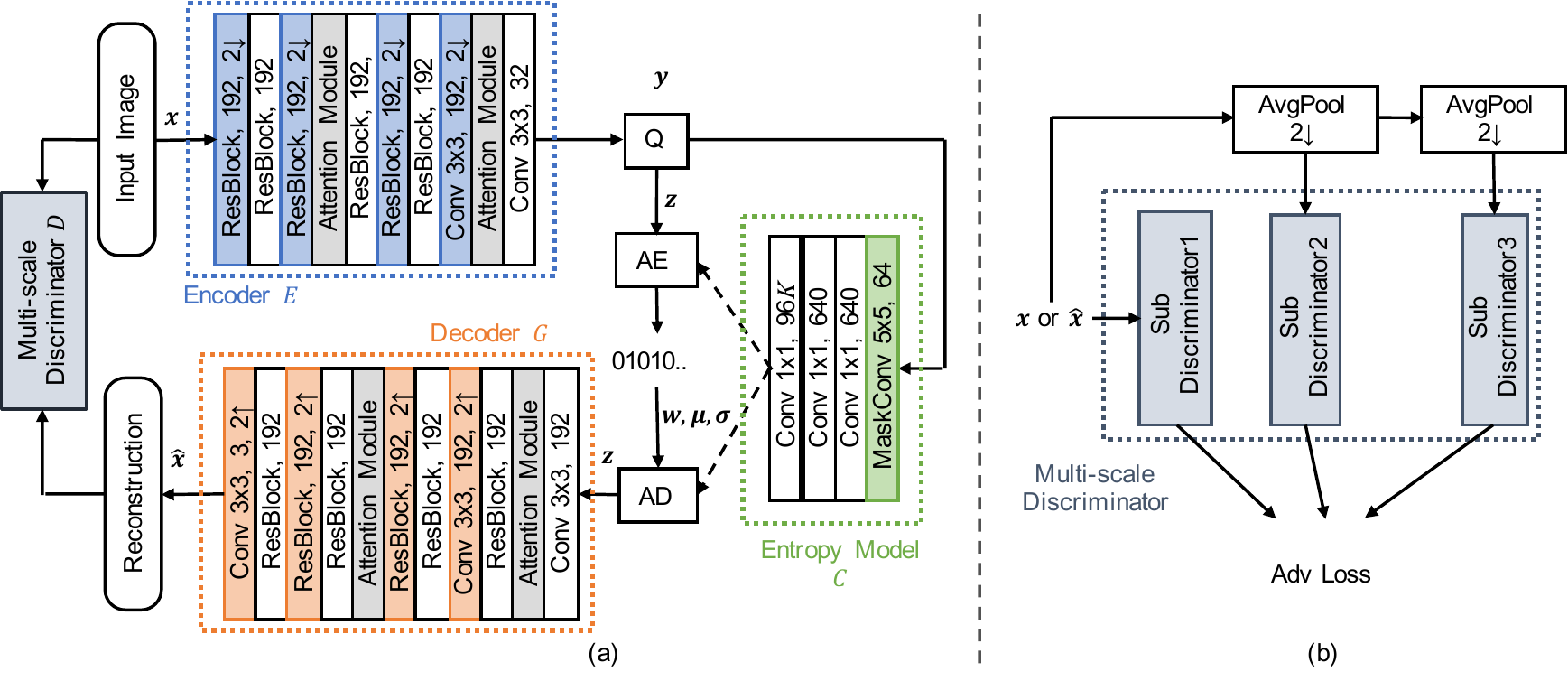}
    \caption{Network architecture. (a) Overall architecture. (b) The multi-scale discriminator architecture. It has three sub-discriminators.}
    \label{fig:network_all}
\end{figure*}
As shown in Fig.~\ref{fig:network_all}, our model has five components: an encoder $E$, a quantizer $Q$, a decoder $G$, a context-based entropy model $C$, and a discriminator $D$.

The encoder $E$ transforms a real image $\bm{x}$ into a latent code $\bm{y}$.
It consists of six residual blocks, two attention modules, and two convolution layers. We use simplified attention modules to reduce the calculation cost. 

The quantizer $Q$ quantizes $\bm{y}$ and produces the quantized code $\bm{z}$.
Following to \cite{balle17}, we use additive uniform noise $\mathcal{U}(0.5,-0.5)$ for quantization during training, and we use $\rm{ROUND}(\cdot)$ during inference. 
\begin{align}
    \bm{z} = \begin{cases}
        \bm{y} + \mathcal{U} \left( \frac{1}{2},-\frac{1}{2} \right) &\rm{(training)}\\
        \rm{ROUND} (\bm{y}) &\rm{(inference)}
        \end{cases}
\end{align}

The entropy model $C$ consists of one masked convolution layer and three $1 \times 1$ convolution layers, as shown in Fig.~\ref{fig:network_all}~(a). 
The masked convolution layer extracts information from the known subset of $\bm{z}$, therefore, it works as the context. Then, we estimate the parameters of distributions $p(\bm{z})$ through the rest three convolution layers.
Since we use a Gaussian mixture model to represent $p(z_i)$, 
it is defined by using $3K$ parameters: means $\bm{\mu}_i = \{\mu_i^{(1)}, \cdots, \mu_i^{(K)}\}$, standard deviations $\bm{\sigma}_i = \{\sigma_i^{(1)}, \cdots, \sigma_i^{(K)}\}$, and weights $\bm{w}_i=\{w_i^{(1)}, \cdots, w_i^{(K)}\}$.
$K$ denotes the number of mixtures.
Thus, $p(\bm{z})$ is as follows:

\begin{eqnarray}
    p(\bm{z}\,|\,\bm{\theta}_{C}) &=& \prod_{i} \left ( p(z_{i}\,|\,\bm{\theta}_{C})  * \mathcal{U}(-\frac{1}{2}, \frac{1}{2}) \right ) (z_i) \nonumber \\
    p(z_{i}\,|\,\bm{\theta}_{C}) &=& \sum_{k=1}^{K} w^{(k)}_{i} \mathcal{N}(\mu^{(k)}_{i}, \sigma^{2(k)}_{i}) \nonumber \\
    && \mathrm{with}\, \bm{w}_{i},\bm{\mu}_{i},\bm{\sigma}_{i}=C(\bm{z}_{<i};\bm{\theta}_{C}).
\end{eqnarray}
$\bm{\theta}_{C}$ denotes the parameters of $C$, and
$\bm{z}_{<i}$ represents the known subset of $z$ at the $i$-th element, such as $\bm{z}_{<i}=(z_1, \cdots, z_{i-1})$.
Then, we transform $\bm{z}$ into a bit stream through the entropy coding.
In this way, $C$ can control the bit allocation. It learns to reduce bits of simple regions and allocate more bits to complex regions over the images.

The decoder $G$ generates a reconstrcution $\hat{\bm{x}}$ from $\bm{z}$. It is a reverse structure of the encoder. We adopt subpixel convolutions for upsampling.

The discriminator $D$ is trained to distinguish $\bm{x}$ from $\hat{\bm{x}}$. Our discriminator is based on the multi-scale discriminator, which was proposed in \cite{pix2pixhd}. The multi-scale discriminator has three sub-discriminators. Each sub-discriminator has an identical architecture but works at different scales as shown in Fig.~\ref{fig:network_all}(b).

\subsection{Two-stage Training}

\begin{figure}[t]
    \centering
    \includegraphics[width=\linewidth]{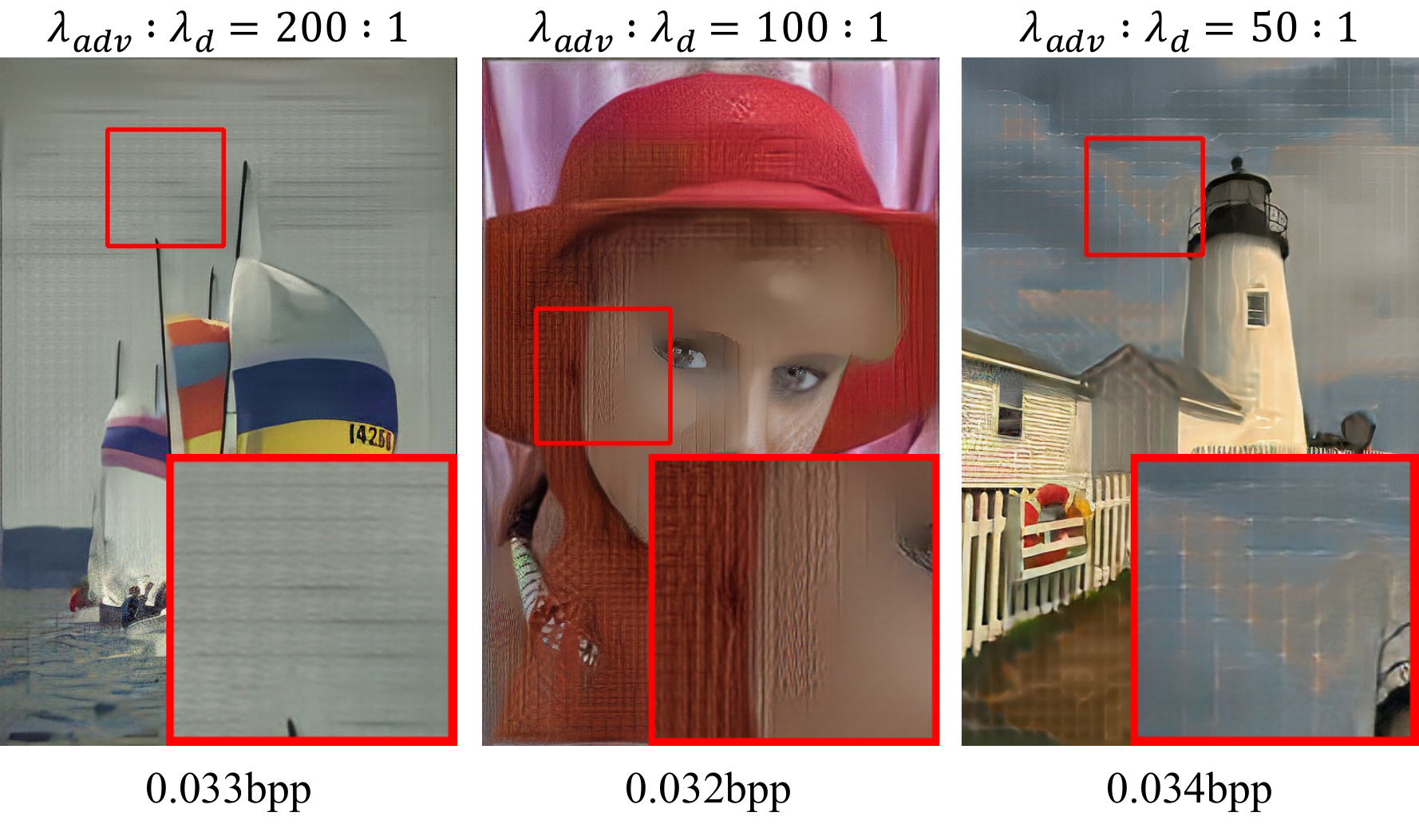}
    \caption{The reconstructions of end-to-end trained models. We tried 3 patterns of the coefficient in (\ref{eq:loss}). We used MSE as $\mathcal{L}_d$. All of the reconstructions suffer from undesirable noise.}
    \label{fig:e2e}
\end{figure}

Our goal is to minimize the loss function $\mathcal{L}$ in Eq.~\ref{eq:loss}. 
There are three trade-off terms: rate $\mathcal{L}_R$, distortion $\mathcal{L}_d$, and perception $\mathcal{L}_{adv}$. The details are described in section \ref{sec:method:loss}.
\begin{equation}
\label{eq:loss}
    \mathcal{L} = \lambda_R \mathcal{L}_R(\bm{z}) + 
    \lambda_d \mathcal{L}_d(\bm{x},\hat{\bm{x}}) + 
    \lambda_{adv} \mathcal{L}_{adv}
\end{equation}

We trained the model to optimize (\ref{eq:loss}) in end-to-end manner. 
However, (\ref{eq:loss}) is hard for minimizing.
As shown in Fig.~\ref{fig:e2e}, the reconstructions suffer from noises, blur, or artifacts and perceptual qualities are quite low.
According to \cite{rdp_tradeoff}, there is a triple trade-off between rate, distortion, and preception. Hence, the end-to-end training failed because the model had to optimize three trade-off terms at the same time.

To tackle this problem, we utilize a two stage training \cite{lee_clic}. Fig.~\ref{fig:overall} shows the training scheme. In the first stage, we train the encoder $E$, decoder $G^1$, and entropy model $C$ to optimize rate-distortion trade-off, that is, we use no adversarial loss ($\lambda_{adv}=0$). 
In the second stage, we initialize the parameters of the decoder $G^2$ with $G^1$ and fine-tune only $G^2$ to optimize distortion-perception trade-off.
Since the parameters of the encoder and entropy model are fixed, $\bm{z}$ and $p(\bm{z})$ do not change through the second stage. Thus we can set $\lambda_{R}$ to 0.

This training method has three advantages.
First, the two-stage training avoids optimizing triple trade-off directly. The model is trained to optimize double trade-off twice and it is effective for a stable training.
Second, after the first stage, the decoder can reconstruct plausible images. Therefore, the discriminator cannot distinguish between real and fake easily even at the very beginning of training. It can lead to a balanced adversarial training.
Third, we can control the balance between perceptual quality and fidelity of reconstructions through network interpolation \cite{esrgan} without re-training the model. Therefore, the two-stage training is also effective to tackle the problem of noise or artifact. 
We will describe the details in the next subsection.

\subsection{Network Interpolation}
It is desirable that reconstructions are realistic and similar to their original images. However, there is a trade-off between distortion and perceptual quality. Thus, improving perceptual quality using adversarial training leads to a high distortion. The reconstructions often contain undesirable noises or artifacts and the impression can be changed. It is a serious problem in image compression.

To tackle this problem, we use the network interpolation \cite{esrgan}. After two-stage training, we have two decoders: $G^1$ and $G^2$ as shown in Fig.~\ref{fig:overall}. The outputs of $G^1$ are high fidelity but blurry because we use no adversarial loss at the first stage. Whereas, the outputs of $G^2$ are good perceptual quality but noisy. Since these decoders share the same encoder and entropy model, we can reconstruct two images from the same latent code. Consequently, we can obtain more desirable images by interpolating the two decoders.

\begin{figure}[t] \centering
    \includegraphics[width=\linewidth]{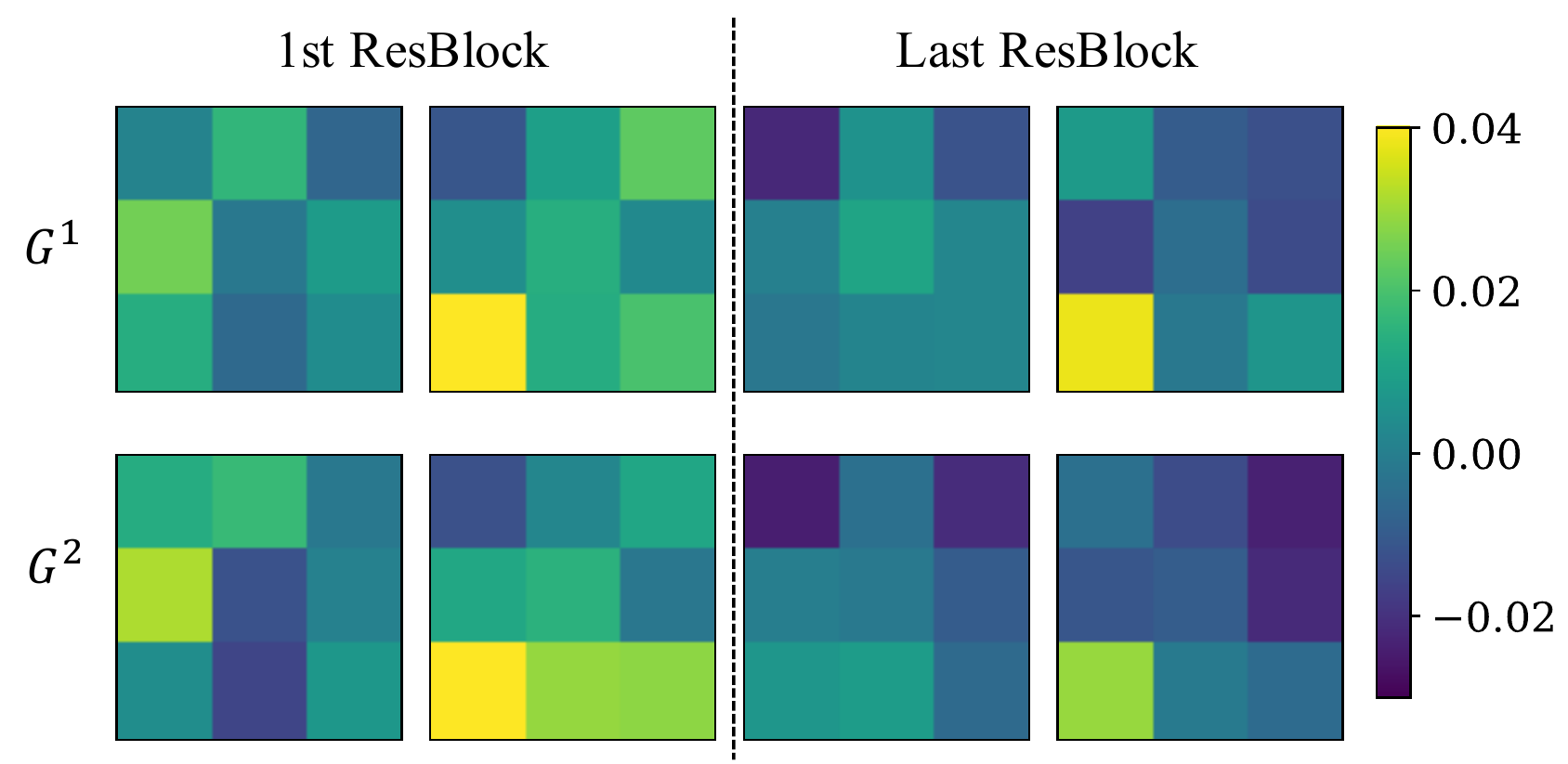}
    \caption{The visualization of the examples of convolution kernels in the decoders. The top and bottom row show the kernel before and after fine-tuning, respectively. We pick up some kernels from the first convolution layer in the first and the last ResBlock in the both decoders.}
    \label{fig:decoder_kernel}
\end{figure}

There are two interpolation methods. The first method is a network interpolation. Fig.~\ref{fig:decoder_kernel} shows convolution kernels of the two decoders. It shows that though the scales of the parameters are changed after fine-tuning, the patterns of the kernels are not changed significantly. Therefore, we interpolate all the corresponding parameters of the two decoders by (\ref{net_interp}).
\begin{equation}
\label{net_interp}
\bm{\theta}_{G'} = (1-\alpha) \bm{\theta}_{G^1} + \alpha \bm{\theta}_{G^2} 
\end{equation}
$\bm{\theta}_{G'}$, $\bm{\theta}_{G^1}$, and $\bm{\theta}_{G^2}$ are the parameter of the interpolated decoder $G'$, $G^1$, and $G^2$, respectively.
$\alpha \in [0, 1]$ is the interpolation parameter. By adjusting $\alpha$, we can continuously control the distortion-perception trade-off without re-training the model.

Another method is an image interpolation. We produce image $\hat{x}'$ by merging reconstructions of the two decoders pixel by pixel.
\begin{equation}
\label{img_interp}
    \hat{\bm{x}}'= (1-\alpha) \hat{\bm{x}}^1 + \alpha \hat{\bm{x}}^2
\end{equation}
$\hat{\bm{x}}'$, $\hat{\bm{x}}^1$, and $\hat{\bm{x}}^2$ are the interpolated reconstruction, the reconstruction of $G^1$ and $G^2$, respectively.

\subsection{Loss function} \label{sec:method:loss}
We define two loss functions $\mathcal{L}_{1st}$ and $\mathcal{L}_{2nd}$ for the first and second stages, respectively.

For the first stage, we define the loss function $\mathcal{L}_{1st}$ as (\ref{eq:loss:1st}), where $N$ is the number of pixels of the image $x$. We use the rate term $\mathcal{L}_R$ and the distortion term $\mathcal{L}_d$. $\mathcal{L}_R$ represents an estimated bitrate of $\bm{z}$. 
In this stage, we use mean squared error (MSE) between $\bm{x}$ and $\hat{\bm{x}}$ as the distortion loss.
\begin{eqnarray}
    \mathcal{L}_{1st} &=& \min_{E,G^1,C} 
    \mathbb{E} [ \mathcal{L}_R(\bm{z}) ] + 
    \lambda_d^{(1)} \mathbb{E} [\mathcal{L}_d(\bm{x}, \hat{\bm{x}})] \label{eq:loss:1st} \\
    \mathcal{L}_R(\bm{z}) &=& - \frac{1}{N} \log_2 p(\bm{z} | \bm{\theta}_C) \label{eq:loss:rate}
\end{eqnarray}
 
For the second-stage, the loss function is defined in (\ref{eq:loss:2nd}). We use MSE $\mathcal{L}_d$, adversarial loss $\mathcal{L}_{adv}$, and VGG loss $\mathcal{L}_{\rm{vgg}}$. For the adversarial training, we adopt Least-Squares GAN \cite{lsgan}. 
\begin{eqnarray}
    \mathcal{L}_{2nd} &=& \min_{G^2} \lambda_d^{(2)} 
    \mathbb{E} [\mathcal{L}_d(\bm{x}, \hat{\bm{x}})] + 
    \lambda_{adv} \mathcal{L}_{adv}^G + \nonumber \\
    && \hspace{7mm} \lambda_{\rm{vgg}} \mathbb{E} [\mathcal{L}_{\rm{vgg}}(\bm{x}, \hat{\bm{x}})] 
    \label{eq:loss:2nd}\\
    \mathcal{L}_{adv}^G &=& \min_{G^2} \mathbb{E} [(D(\hat{\bm{x}})-1)^2]\\
    \mathcal{L}_{adv}^D &=& \min_D \mathbb{E} [(D(\hat{\bm{x}}))^2] + \mathbb{E} [(D(x)-1)^2]
\end{eqnarray}
The VGG loss $\mathcal{L}_{\rm{vgg}}$ is the L1 distance between the feature maps of VGG-19 \cite{vgg19}. As in \cite{esrgan}, we use the feature map obtained in the fourth convolution before the fifth max-pooling layer. It helps the model to reconstruct sharp image.


\section{Experiments}

\begin{figure*}[t] \centering
    \includegraphics[width=\linewidth]{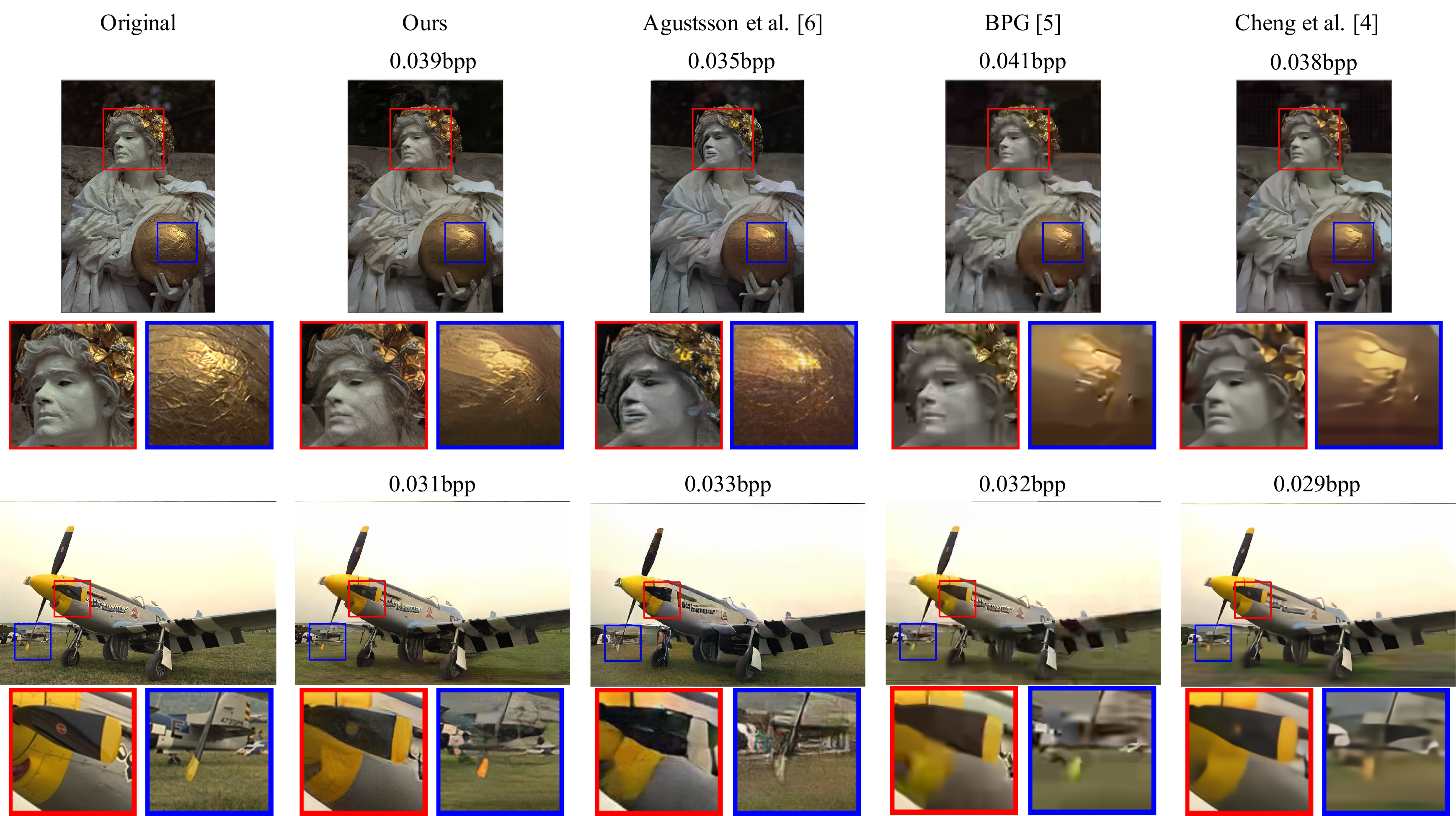}
    \caption{Visual comparison of the original image, reconstructions of ours, Agustsson et al\cite{agustsson}, BPG\cite{bpg} and Cheng et al.\cite{gmm}}
    \label{fig:others_compare}
\end{figure*}

\subsection{Experiments Details}
For training, we used $256 \times 256$ patches extracted from 118287 images in COCO dataset \cite{coco}. For evaluation, we used Kodak PhotoCD image dataset \cite{kodak}. We set the batch size to 8. We trained the whole model for 500k iterations in the first stage and fine-tuned only the decoder for 300k iterations in the second stage. The learning rate is set to $2\times 10^{-5}$ and halved at [150k, 300k] iterations. We applied Adam optimizer \cite{adam}. 
The number of mixture $K$ is set to 3.\par
For the first training, $\lambda_d^{(1)}$ depends on our ideal bitrates. We trained four models with different compression rates. For the second training, we minimize the loss in (\ref{eq:loss:2nd}) with $\lambda_d^{(2)}=0.01$, $\lambda_{adv}=1$, and $\lambda_{\rm{VGG}}=20$. Note that, when we calculate MSE, the ranges of $x$ and $\hat{x}$ are $[0,255]$. We use network interpolation to reduce undesirable noises. We set the interpolation parameter $\alpha$ to $0.8$.

\subsection{Qualitative Result}
We compare our method on the Kodak dataset with the state-of-the-art engineered compression codec BPG \cite{bpg} (in the 4:2:0 chroma format), the GAN-based image compression method Agustsson et al. \cite{agustsson}, and the PSNR-oriented compression method Cheng et al.\cite{gmm} at extreme low bitrate. Fig.~\ref{fig:others_compare} shows the original Kodak images, the reconstructions, and their bitrates. Though the reconstructions of Agustsson et al. are sharp, they have undesirable artifacts and the impressions of the images are changed.
For instance, the face of the statue looks unnatural in the top image, and the color of aircraft's propeller is changed in the bottom image.
BPG suffers from block noises and blur. Cheng et al. does not have unpleased noises, however, the reconstructions are blurry.  Our reconstructions do not suffer from artifacts and blur. They are natural and high-fidelity.

\subsection{User Study}

\begin{table}[t]
    \caption{The information of images which we used in our user study.}
    \begin{center}
    \begin{tabular}{|c|c|c|c|}
    \hline
    \textbf{}&\textbf{}&\multicolumn{2}{|c|}{\textbf{Average bitrate [bpp]}} \\
    \cline{3-4} 
    \textbf{model} & \textbf{pairs} & \textbf{Ours}& \textbf{Agustsson et al.}\\
    \hline
    0.03bpp & 20 & 0.0332 & 0.0333  \\
    0.06bpp & 11 & 0.0644 & 0.0678  \\
    total & 31 & 0.0445 & 0.0443  \\
    \hline
    \end{tabular}
    \label{tab:user_study_table}
    \end{center}
\end{table}

\begin{figure}[t]
    \centering
    \includegraphics[width=\linewidth]{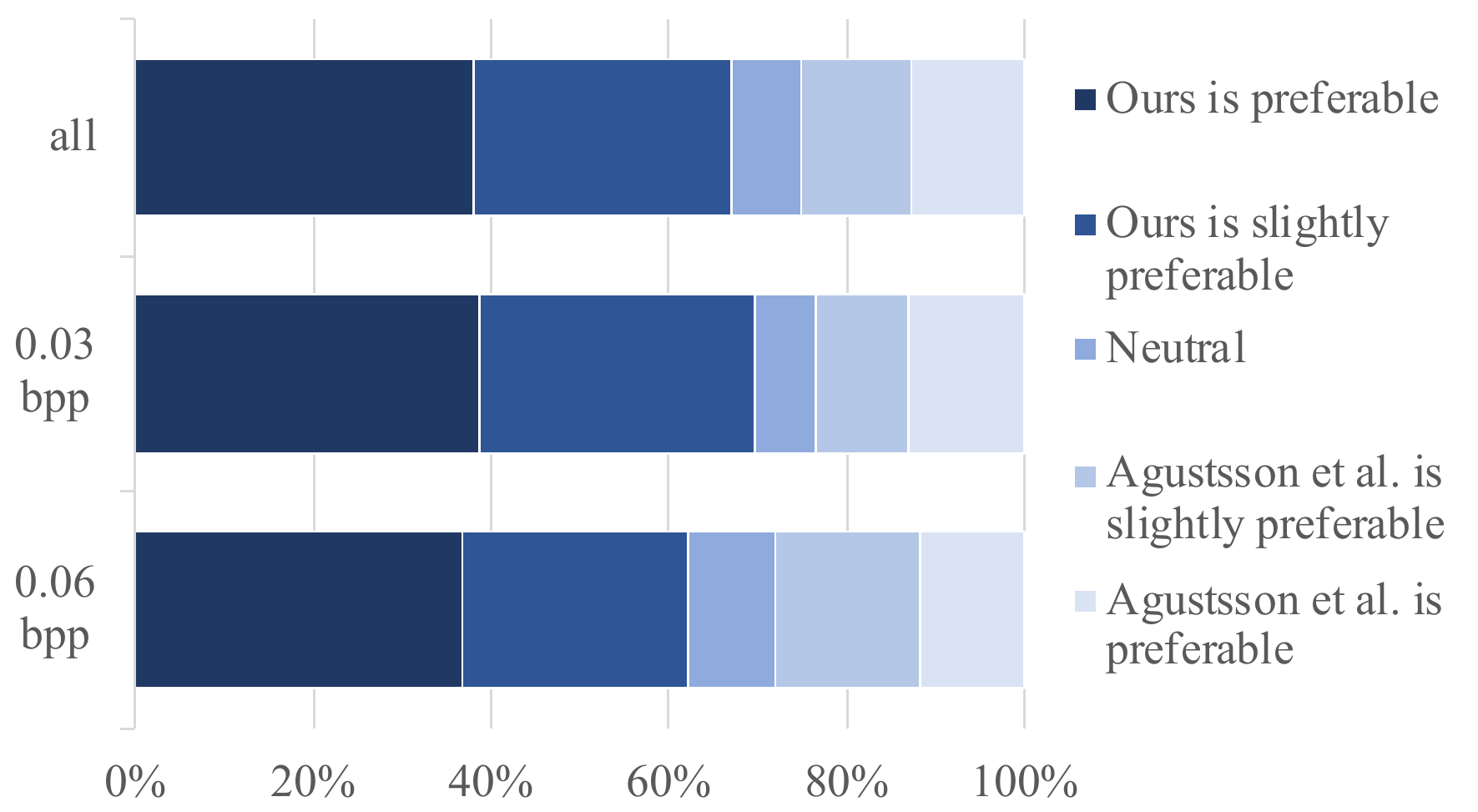}
    \caption{User study results on Kodak dataset.}
    \label{fig:user_study_all}
\end{figure}

\begin{figure}[t]
    \centering
    \includegraphics[width=\linewidth]{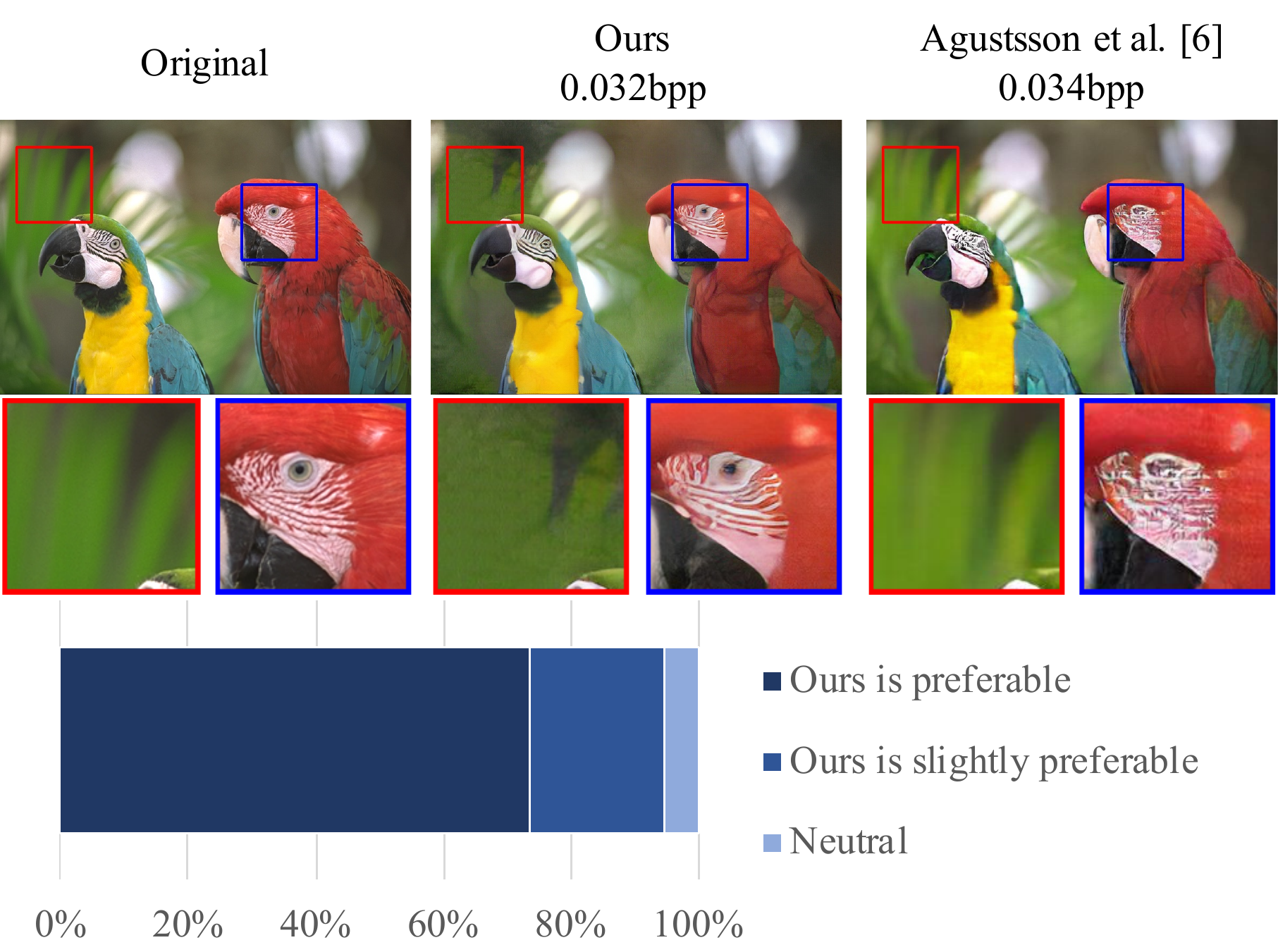}
    \caption{An example of user study results. Over 90\% users answered "Ours is preferable" or "Ours is slightly preferable".}
    \label{fig:user_study_kodim23}
\end{figure}

We performed a user study to compare our method and Agustsson et al.\cite{agustsson}. \cite{agustsson} has two models: 0.03bpp model and 0.06bpp model. 
We chose 31 pairs of the reconstructions of ours and \cite{agustsson} so that the difference of bitrates between them is less than 0.005bpp and 0.01bpp for the 0.03bpp model and the 0.06bpp model, respectively.
The summary of the pairs are shown in Table~\ref{tab:user_study_table}. The average bitrates of ours and \cite{agustsson} 0.03bpp model are 0.0332bpp and 0.0333bpp, respectively, and ours and \cite{agustsson} 0.06bpp model are 0.0644bpp and 0.0678bpp, respectively. We showed the users the original image and two reconstructions. The users were requested to evaluate which reconstruction is preferable as compressed image on a scale of 1 to 5. We asked 19 users.\par
As shown in Fig.~\ref{fig:user_study_all}, our method outperforms Agustsson et al. in both 0.03bpp model and 0.06bpp model. More than 60\% of the answers are "Ours is preferable" or "Ours is slightly preferable". Especially, our method achieves high score against \cite{agustsson} 0.03bpp model. This indicates that our method provides good performances even at extreme low bitrates (around 0.03bpp). \par
Fig.~\ref{fig:user_study_kodim23} shows an example which our reconstruction is highly evaluated. Over 90\% users answered "Ours is preferable" or "Ours is slightly preferable" to this image.
When we focus on the background of the image, the reconstruction of Agustsson et al. looks better. However, the faces of the birds collapse and the reconstruction looks unnatural. In contrast, our method succeeds in reconstructing the faces because it can allocate more bits to complex regions. Thus, a lot of users value our method highly.


\subsection{Network Interpolation}

\begin{figure}[t]
    \centering
    \includegraphics[width=\linewidth]{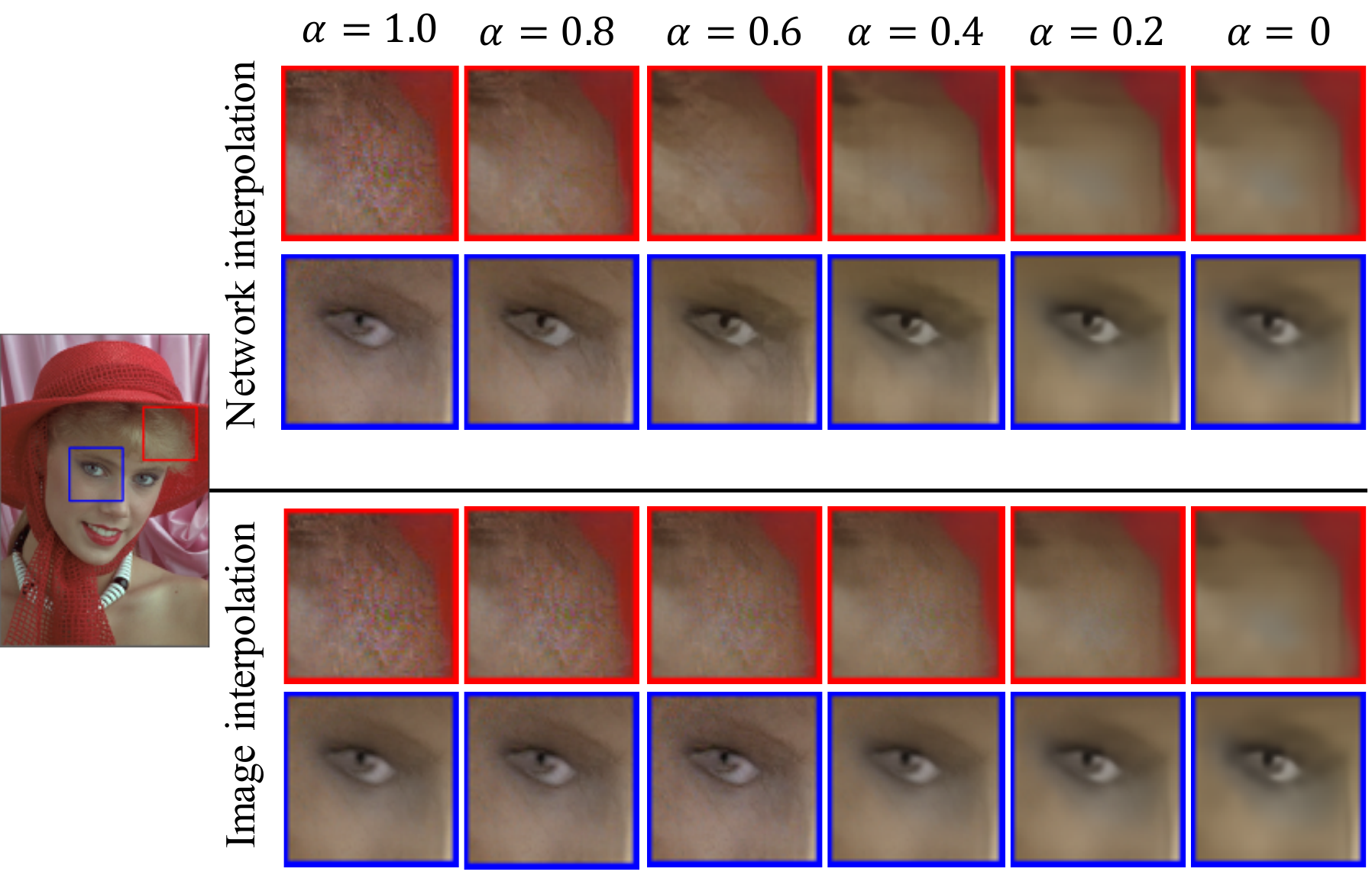}
    \caption{Visual comparison between network interpolation and image interpolation.}
    \label{fig:interpolation}
\end{figure}

We compare the two interpolation methods: network interpolation (\ref{net_interp}) and image interpolation (\ref{img_interp}). We change the interpolation parameter $\alpha$ from 0 to 1 with an interval of $0.2$. As shown in Fig.~\ref{fig:interpolation}, in both methods, when we decrease $\alpha$, we can reduce noises but the reconstructions become blurry. In contrast, when we increase $\alpha$, the reconstructions become sharper but they suffer from artifacts or noises. We can balance noise and blur without re-training the model by adjusting $\alpha$.\par
Though both methods can reduce noises, the network interpolation works better than the image interpolation in image compression as well as super-resoluton \cite{esrgan}.
In Fig.~\ref{fig:interpolation}, though the network interpolation reduces the noise at $\alpha=0.8$, the image interpolation suffers from noise even at $\alpha=0.6$.
The network interpolation can reduce the noise while it keeps the image sharp. In contrast, the image interpolation reduces noises as well as edges. 

\subsection{Ablation Study}

\begin{figure}[t]
    \centering
    \includegraphics[width=\linewidth]{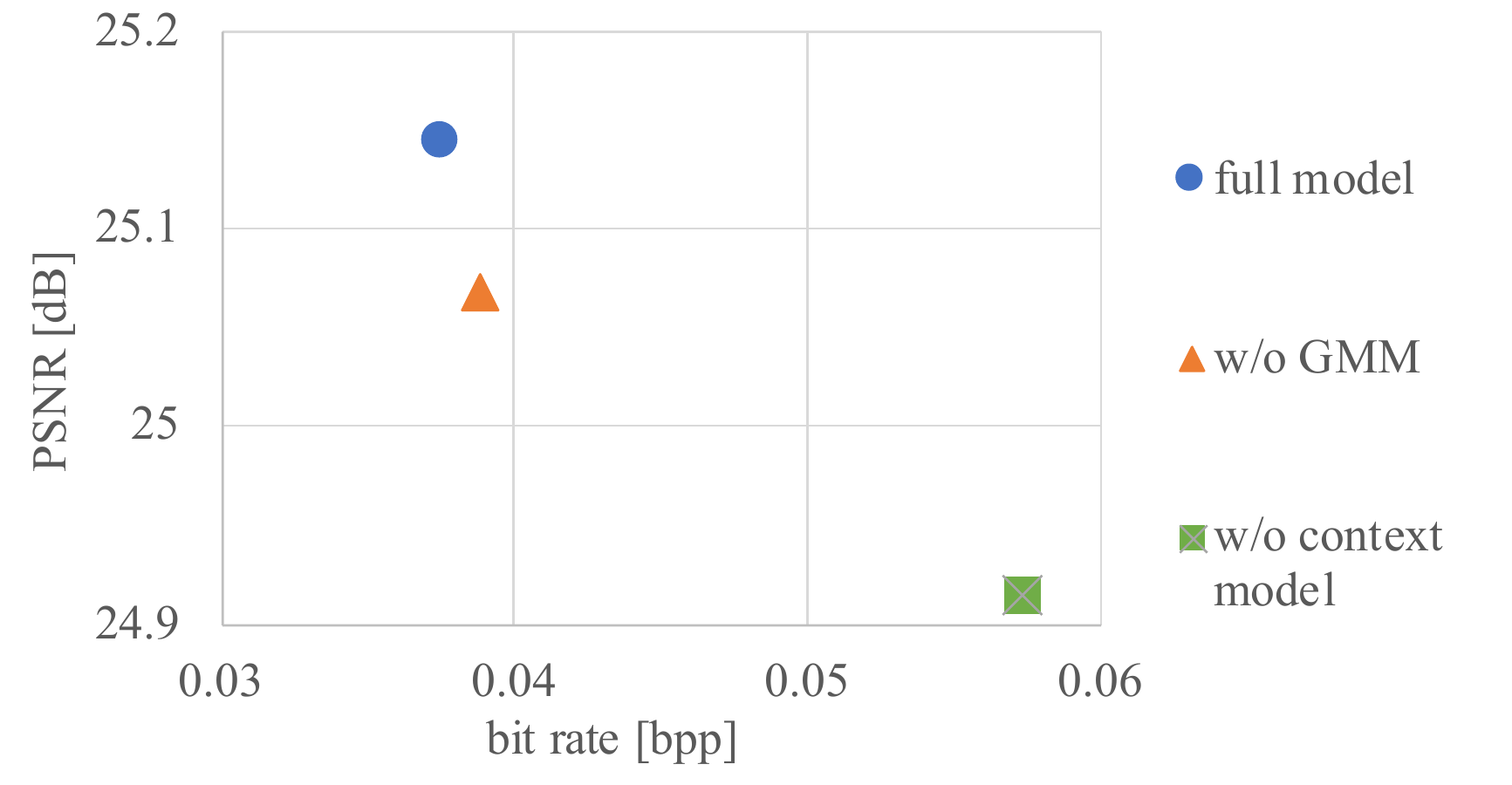}
    \caption{Quantitative comparison among our full model, w/o GMM and w/o context model. PSNR and bitrates are calculated on Kodak dataset.}
    \label{fig:ablation_graph}
\end{figure}

\begin{figure*}[t]
    \centering
    \includegraphics[width=\linewidth]{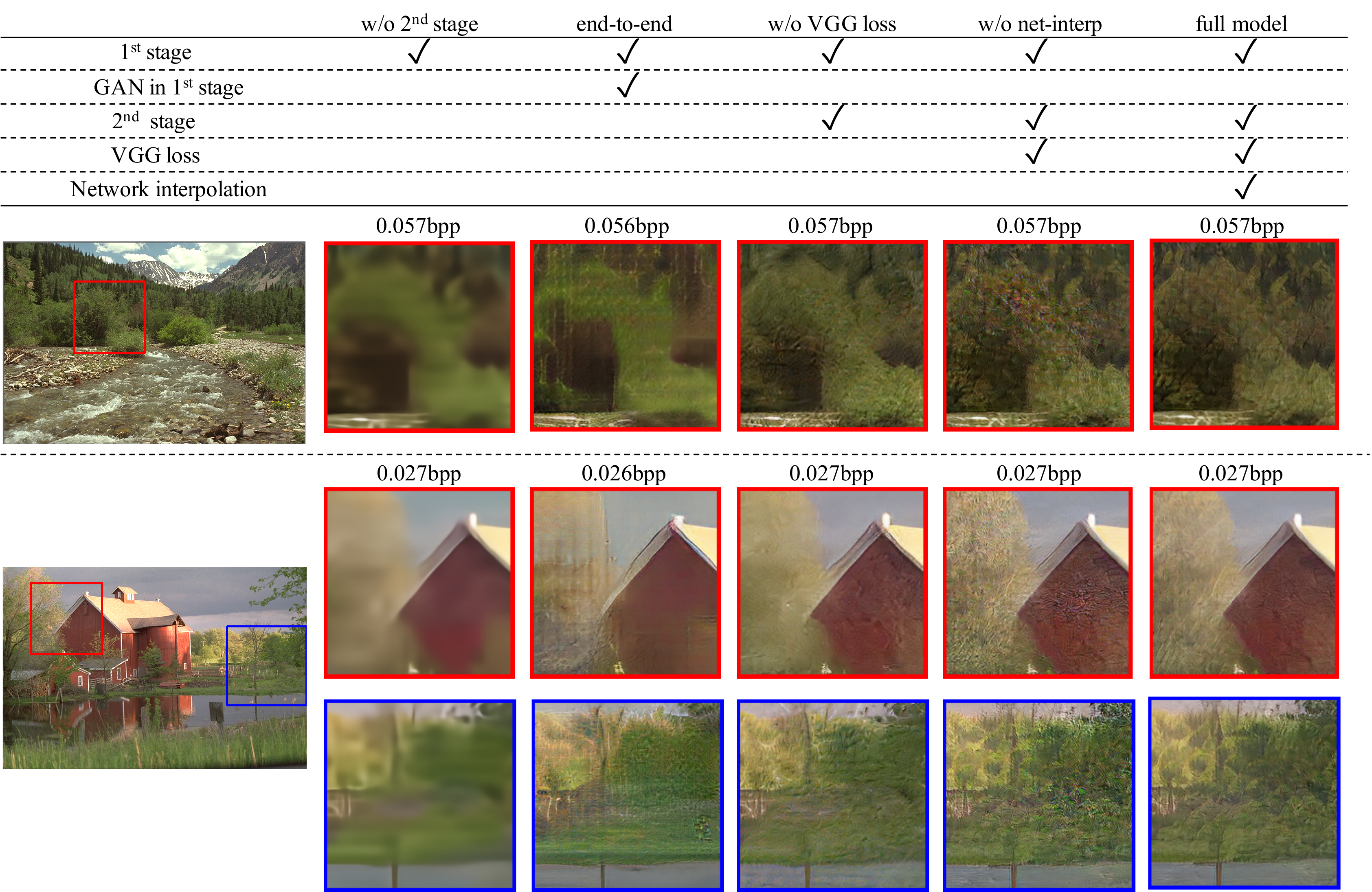}
    \caption{Visual comparison among original images, w/o 2nd stage, end-to-end, w/o VGG loss, w/o network interpolation and our full model on Kodak dataset.}
    \label{fig:ablation_compare}
\end{figure*}

First, we analyze the effects of the entropy model which utilizes the Gaussian mixture model and the context model. We trained two models: w/o GMM and w/o context.
We will explain the details.
\subsubsection*{w/o GMM}
We use a single Gaussian model instead of a Gaussian mixture model in the entropy model. Therefore, the entropy model estimates mean and standard deviation parameters of the distribution of quantized code.
\subsubsection*{w/o context model}
We use a simple entropy model which has no layers. We assume the distribution $p(\bm{z})$ follows Gaussian models which have a learnable standard deviation. 

Fig.~\ref{fig:ablation_graph} shows the average bitrate and PSNR of Kodak dataset in the first stage. Note that since the entropy model does not affect the second stage, we evaluate them on the first stage. Our full model outperforms w/o GMM and w/o context. It indicates that the Gaussian mixture model is effective even below 0.05bpp.\par
Next, we evaluate the training scheme and loss functions. We trained four models for comparison. Detailed settings are as follows.
\subsubsection*{w/o 2nd stage}
We do not fine-tune the model. The model is trained to optimize rate-distortoin trade-off and adversarial training is not used in this setting.
\subsubsection*{end-to-end}
We train this model in end-to-end manner. The loss function is:
\begin{eqnarray}
    \mathcal{L}_{E2E} = \lambda_{R} \mathcal{L}_{R} (\hat{y}) + \lambda_d 
    \mathbb{E} [\mathcal{L}_d(\bm{x}, \hat{\bm{x}})] + 
    \lambda_{adv} \mathcal{L}_{adv}^G + \nonumber \\
    \lambda_{\rm{vgg}} \mathbb{E} [\mathcal{L}_{\rm{vgg}}(\bm{x}, \hat{\bm{x}})].
\end{eqnarray}
We set $\lambda_{R}$, $\lambda_d$, $\lambda_{adv}$, and $\lambda_{\rm{vgg}}$ to 2.4, 0.01, 1, and 20, respectively. 
\subsubsection*{w/o VGG loss}
This model is trained using two-stage training. We use no VGG loss in the second stage.
\par
\subsubsection*{w/o net-interp}
This model is trained in the same way as our full model, however, we do not use network interpolation.\par

In Fig.~\ref{fig:ablation_compare}, we provide comparison of the baselines and our full model. The reconstructions of w/o fine-tuning are blurry. Since this model is trained to optimize rate-distortion trade-off, the perceptual quality become worse. The end-to-end trained model suffers from unnatural noise and artifacts. Though the reconstructions are not blurry, perceptual qualities are poor. The w/o VGG loss model reconstructs relatively plausible images, however, it fails to generate fine texture. For instance, in the bottom image in Fig.~\ref{fig:ablation_compare}, the leaves of the tree are blurry. In contrast, the w/o net-interp generates fine details. It shows that VGG loss helps the model to add finer textures to reconstructions. However, the reconstructions contain noises. Our full model can reconstruct visually pleasing images. Network interpolation reduces noise while maintaining fine details. 


\section{Conclusion}
We proposed GAN-based extreme image compression method. Our method utilizes Gaussian mixture model in entropy model and we showed that it works well even at very low bitrates. We adopt the two-stage training and the network interpolation to tackle the two main problems of GAN-based compression methods: unstable training and undesirable noises or artifacts. Our reconstructions are perceptually high quality and high fidelity. Moreover, our user study shows the proposed method outperforms Agustsson et al.\cite{agustsson} in terms of the quality as compressed images.

\bibliographystyle{IEEEtran}
%
\bibliography{IEEEabrv, bibliography} 

\begin{thebibliography}{10}
\providecommand{\url}[1]{#1}
\csname url@samestyle\endcsname
\providecommand{\newblock}{\relax}
\providecommand{\bibinfo}[2]{#2}
\providecommand{\BIBentrySTDinterwordspacing}{\spaceskip=0pt\relax}
\providecommand{\BIBentryALTinterwordstretchfactor}{4}
\providecommand{\BIBentryALTinterwordspacing}{\spaceskip=\fontdimen2\font plus
\BIBentryALTinterwordstretchfactor\fontdimen3\font minus
  \fontdimen4\font\relax}
\providecommand{\BIBforeignlanguage}[2]{{%
\expandafter\ifx\csname l@#1\endcsname\relax
\typeout{** WARNING: IEEEtran.bst: No hyphenation pattern has been}%
\typeout{** loaded for the language `#1'. Using the pattern for}%
\typeout{** the default language instead.}%
\else
\language=\csname l@#1\endcsname
\fi
#2}}
\providecommand{\BIBdecl}{\relax}
\BIBdecl

\bibitem{balle17}
J.~Ball{\'{e}}, V.~Laparra, and E.~P. Simoncelli, ``End-to-end optimized image
  compression,'' in \emph{International Conference on Learning Representations,
  {ICLR}}, 2017.

\bibitem{balle18}
J.~Ball{\'{e}}, D.~Minnen, S.~Singh, S.~J. Hwang, and N.~Johnston,
  ``Variational image compression with a scale hyperprior,'' in
  \emph{International Conference on Learning Representations, {ICLR}}, 2018.

\bibitem{minnen}
D.~Minnen, J.~Ball\'{e}, and G.~Toderici, ``Joint autoregressive and
  hierarchical priors for learned image compression,'' in \emph{Advances in
  Neural Information Processing Systems}, 2018, pp. 10\,771--10\,780.

\bibitem{gmm}
Z.~Cheng, H.~Sun, M.~Takeuchi, and J.~Katto, ``Learned image compression with
  discretized gaussian mixture likelihoods and attention modules,'' in
  \emph{IEEE/CVF Conference on Computer Vision and Pattern Recognition (CVPR)},
  June 2020.

\bibitem{bpg}
\BIBentryALTinterwordspacing
F.~Bellard, ``Bpg image format.'' [Online]. Available:
  \url{https://bellard.org/bpg/}
\BIBentrySTDinterwordspacing

\bibitem{agustsson}
E.~Agustsson, M.~Tschannen, F.~Mentzer, R.~Timofte, and L.~V. Gool,
  ``Generative adversarial networks for extreme learned image compression,'' in
  \emph{The IEEE International Conference on Computer Vision (ICCV)}, October
  2019.

\bibitem{rippel}
O.~Rippel and L.~D. Bourdev, ``Real-time adaptive image compression,'' in
  \emph{International Conference on Machine Learning, {ICML}}, ser. Proceedings
  of Machine Learning Research, vol.~70.\hskip 1em plus 0.5em minus 0.4em\relax
  {PMLR}, 2017, pp. 2922--2930.

\bibitem{a_gan_based}
L.~Wu, K.~Huang, and H.~Shen, ``A gan-based tunable image compression system,''
  in \emph{IEEE Winter Conference on Applications of Computer Vision (WACV)},
  2020, pp. 2323--2331.

\bibitem{lee_clic}
J.~Lee, D.~Kim, Y.~Kim, H.~Kwon, J.~Kim, and T.~Lee, ``A training method for
  image compression networks to improve perceptual quality of
  reconstructions,'' in \emph{The IEEE/CVF Conference on Computer Vision and
  Pattern Recognition (CVPR) Workshops}, June 2020.

\bibitem{esrgan}
X.~Wang, K.~Yu, S.~Wu, J.~Gu, Y.~Liu, C.~Dong, Y.~Qiao, and C.~C. Loy,
  ``Esrgan: Enhanced super-resolution generative adversarial networks,'' in
  \emph{The European Conference on Computer Vision Workshops (ECCVW)},
  September 2018.

\bibitem{GAN}
I.~Goodfellow, J.~Pouget-Abadie, M.~Mirza, B.~Xu, D.~Warde-Farley, S.~Ozair,
  A.~Courville, and Y.~Bengio, ``Generative adversarial nets,'' in
  \emph{Advances in Neural Information Processing Systems}, 2014, pp.
  2672--2680.

\bibitem{deblur_gan}
O.~Kupyn, V.~Budzan, M.~Mykhailych, D.~Mishkin, and J.~Matas, ``Deblurgan:
  Blind motion deblurring using conditional adversarial networks,'' \emph{ArXiv
  e-prints}, 2017.

\bibitem{deepfillv2}
J.~Yu, Z.~Lin, J.~Yang, X.~Shen, X.~Lu, and T.~S. Huang, ``Free-form image
  inpainting with gated convolution,'' in \emph{IEEE International Conference
  on Computer Vision (ICCV)}, 2019, pp. 4471--4480.

\bibitem{conditional_probability}
F.~Mentzer, E.~Agustsson, M.~Tschannen, R.~Timofte, and L.~V. Gool,
  ``Conditional probability models for deep image compression,'' in \emph{The
  IEEE Conference on Computer Vision and Pattern Recognition (CVPR)}, June
  2018.

\bibitem{ca_entropy}
J.~Lee, S.~Cho, and S.~Beack, ``Context-adaptive entropy model for end-to-end
  optimized image compression,'' in \emph{International Conference on Learning
  Representations, {ICLR}}, 2019.

\bibitem{NLAIC}
T.~Chen, H.~Liu, Z.~Ma, Q.~Shen, X.~Cao, and Y.~Wang, ``Neural image
  compression via non-local attention optimization and improved context
  modeling,'' 2019.

\bibitem{pixelcnn}
A.~van~den Oord, N.~Kalchbrenner, L.~Espeholt, K.~Kavukcuoglu, O.~Vinyals, and
  A.~Graves, ``Conditional image generation with pixelcnn decoders,'' in
  \emph{Advances in Neural Information Processing Systems}.\hskip 1em plus
  0.5em minus 0.4em\relax Curran Associates, Inc., 2016, pp. 4790--4798.

\bibitem{multi_scale_comp}
K.~Nakanishi, S.~Maeda, T.~Miyato, and D.~Okanohara, ``Neural multi-scale image
  compression,'' \emph{Lecture Notes in Computer Science}, p. 718–732, 2019.

\bibitem{generative_model_dp}
M.~Tschannen, E.~Agustsson, and M.~Lucic, ``Deep generative models for
  distribution-preserving lossy compression,'' in \emph{Advances in Neural
  Information Processing Systems}.\hskip 1em plus 0.5em minus 0.4em\relax
  Curran Associates, Inc., 2018, pp. 5929--5940.

\bibitem{pix2pixhd}
T.~Wang, M.~Liu, J.~Zhu, A.~Tao, J.~Kautz, and B.~Catanzaro, ``High-resolution
  image synthesis and semantic manipulation with conditional gans,'' in
  \emph{The IEEE Conference on Computer Vision and Pattern Recognition (CVPR)},
  June 2018.

\bibitem{rdp_tradeoff}
Y.~Blau and T.~Michaeli, ``Rethinking lossy compression: The
  rate-distortion-perception tradeoff,'' in \emph{International Conference on
  Machine Learning, {ICML}}, ser. Proceedings of Machine Learning Research,
  vol.~97.\hskip 1em plus 0.5em minus 0.4em\relax {PMLR}, 2019, pp. 675--685.

\bibitem{lsgan}
X.~Mao, Q.~Li, H.~Xie, R.~Y.~K. Lau, Z.~Wang, and S.~P. Smolley, ``Least
  squares generative adversarial networks,'' \emph{IEEE International
  Conference on Computer Vision (ICCV)}, Oct 2017.

\bibitem{vgg19}
K.Simonyan and A.~Zisserman, ``Very deep convolutional networks for large-scale
  image recognition,'' 2014.

\bibitem{coco}
\BIBentryALTinterwordspacing
T.-Y. Lin, M.~Maire, S.~Belongie, J.~Hays, P.~Perona, D.~Ramanan, P.~Dollár,
  and C.~L. Zitnick, ``Microsoft coco: Common objects in context,''
  \emph{Lecture Notes in Computer Science}, p. 740–755, 2014. [Online].
  Available: \url{http://dx.doi.org/10.1007/978-3-319-10602-1_48}
\BIBentrySTDinterwordspacing

\bibitem{kodak}
\BIBentryALTinterwordspacing
E.~Kodak, ``Kodak photocd dataset.'' [Online]. Available:
  \url{http://r0k.us/graphics/kodak/}
\BIBentrySTDinterwordspacing

\bibitem{adam}
D.~P. Kingma and J.~Ba, ``Adam: {A} method for stochastic optimization,'' in
  \emph{International Conference on Learning Representations, {ICLR}}, 2015.

\end{thebibliography}



\end{document}